\documentclass[12pt]{elsarticle}

\usepackage{amssymb}
\usepackage{epsfig}
\usepackage{epstopdf}
\usepackage[latin1]{inputenc}
\usepackage{amsthm}

\journal{Physica A}

\begin{document}

\begin{frontmatter}



\title{Normal mode analysis of disordered random-matrix ensembles}


\author[ICN,C3]{R. Fossion}
\ead{fossion@nucleares.unam.mx}
\author[ICBI,UAM]{G. Torres-Vargas}
\ead{gamaliel\_torres@uaeh.edu.mx}

\address[ICN]{Instituto de Ciencias Nucleares, Universidad Nacional Aut\'onoma de M\'exico, 04510 CDMX, Mexico}
\address[C3]{Centro de Ciencias de la Complejidad (C3), Universidad Nacional Aut\'onoma de M\'exico, 04510 CDMX, Mexico}
\address[ICBI]{Instituto de Ciencias B\'asicas e Ingenier\'ia, Universidad Aut\'onoma del Estado de Hidalgo, 42184 Hidalgo, Mexico}
\address[UAM]{Posgrado en Ciencias Naturales e Ingenier\'ia, Universidad Aut\'onoma Metropolitana Cuajimalpa, 05348 CDMX, Mexico}

\begin{abstract}
The statistics of random-matrix spectra can be very sensitive to the unfolding procedure that separates global from local properties. In order to avoid the introduction of possible artifacts, recently it has been applied to ergodic ensembles of Random Matrix Theory (RMT) the singular value decomposition (SVD) method, based on normal mode analysis, which characterizes the long-range correlations of the spectral fluctuations in a direct way without performing any unfolding. However, in the case of more general ensembles, the ergodicity property is often broken leading to ambiguities between spectrum-unfolded and ensemble-unfolded fluctuation statistics. Here, we apply SVD to a disordered random-matrix ensemble with tunable nonergodicity, as a mathematical framework to characterize the nonergodicity. We show that ensemble-averaged and individual-spectrum averaged statistics are calculated consistently using the same normal mode basis, and the nonergodicity is explained as a breakdown of this common basis.
\end{abstract}


\end{frontmatter}


\section{Introduction}
\label{Intro}

Random Matrix Theory (RMT) has been very successful in the statistical study of quantum excitation spectra~\cite{meh91}. The standard Gaussian ensembles of RMT are constructed with matrices whose elements are determined independently from a Gaussian distribution~\cite{haa10}. However, this RMT modeling is not completely realistic, and there has been a search for models whose randomness would mimic physical reality closer. For example, many-body systems are effectively governed by one- and two-body forces, while RMT assumes many-body forces between the constituents, so that a stochastic modeling of the one- and two-body interaction would yield a much smaller number of independent random variables than used in RMT~\cite{ben01,asa01}. Hence the interest in sparse matrices~\cite{jac01}, band- or tridiagonal matrices~\cite{mal07,rel08}, specialized models such as the two-body random ensemble (TBRE)~\cite{fre70,flo01}, and the more general $k-$body embedded Gaussian ensembles (EGE)~\cite{mon75}. Other generalizations determine the matrix elements from a stable but non-Gaussian distribution, in particular the L\'{e}vy distribution~\cite{ciz94}. Also, the statistical properties of addition~\cite{guh90} or multiplication~\cite{bir07,boh08,boh11} of random-matrix variables has been investigated. The main features of the new ensembles are correlations among matrix elements~\cite{tos04,ber04}, Gaussian instead of semicircular global eigenvalue densities~\cite{flo01}, breaking of the power-law behavior of the integrated level density fluctuations~\cite{jac01,mal07,rel08} and nonergodicity~\cite{asa01,fre70,flo01}. These new features pose the question whether a more realistic stochastic modeling of many-body systems might yield results which differ from the standard RMT predictions.\\

In this context, one technical step prior to the statistical study of fluctuations of the standard Gaussian and more general RMT ensembles is the \emph{unfolding} procedure, which serves two purposes: (i) to separate the global level density $\overline{\rho}(E)$ from the local fluctuations $\widetilde{\rho}(E)=\rho(E)-\overline{\rho}(E)$ and (ii) to rescale and normalize the fluctuations so that the statistics of different systems can be compared~\cite{meh91,bro81}. In general, the unfolding is not trivial, and the statistical results can be very sensitive to the particular unfolding procedure applied~\cite{gom02,abu12,she18}. If the ensemble under study is \emph{ergodic}, then \emph{spectrum-unfolded} and \emph{ensemble-unfolded} fluctuation statistics are equivalent. The breaking of ergodicity originates an ambiguity in the characterization of the spectral fluctuations because both measures lead to different results (see e.g.~\cite{asa01}).\\


In previous contributions, we applied a data-adaptive and parameter-free method, the singular value decomposition (SVD), to the standard Gaussian ensembles of RMT, such that each spectrum was decomposed exactly as the sum of trend and fluctuation normal modes, which constitute a basis for the whole ensemble~\cite{fos13b,Torres17,Torres18}. The dominant modes (trend modes) are monotonous and describe the global spectral properties, whereas the other modes oscillate and constitute the fluctuations (fluctuation modes). An advantage of this method, is that an ensemble estimate is obtained for the spectral rigidity in terms of the scaling behavior of the normal modes, in a direct way, without performing any unfolding. On the other hand, the scaling of the fluctuations of each spectrum can also be studied individually, which leads to a spectrum estimate of the ergodicity. In the case of ergodic Gaussian ensembles, both estimates are identical, however, when there are doubts about the validity of the ergodicity hypothesis, the results obtained can lead to ambivalent conclusions. In this sense, a first successful application of SVD to the study of nonergodic ensembles was realized in Ref.~\cite{fos15}, where the TBRE ensemble was considered. In the present contribution, we consider a disordered random-matrix model \cite{boh08,boh11}, that allows to fine-tune the intensity of nonergodicity, to study its effect on the spectrum and ensemble estimates, employing the normal mode analysis.

%
%

\begin{figure*}[htb!]
\begin{minipage}{\linewidth}    
        \begin{minipage}{0.5\linewidth}
        \begin{center}
        \includegraphics[width=\linewidth]{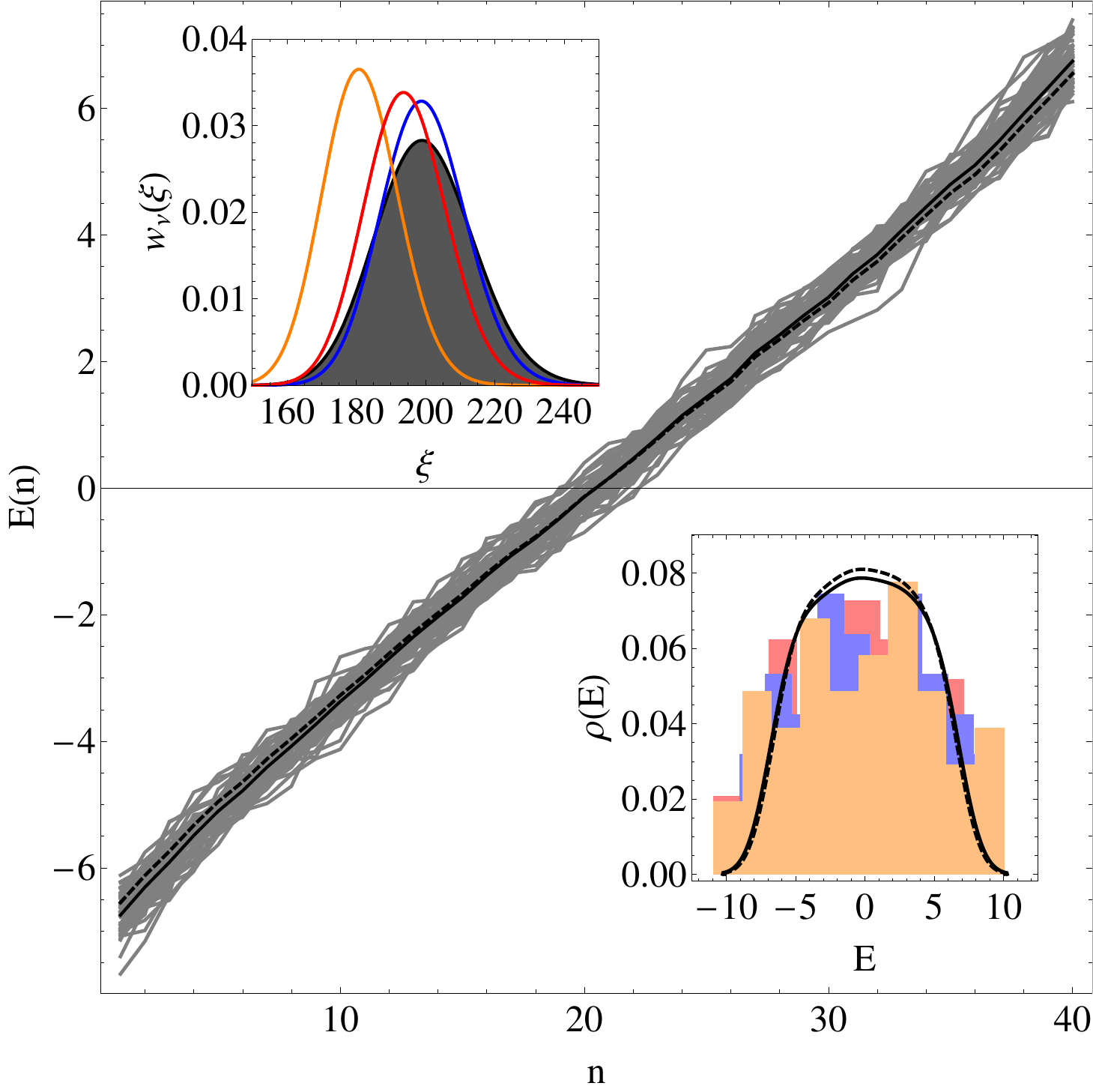}\\
        (b) Ergodic regime ($\overline{\xi}=200$)
        \end{center}
        \end{minipage}
        \begin{minipage}{0.5\linewidth}
        \begin{center}
        \includegraphics[width=\linewidth]{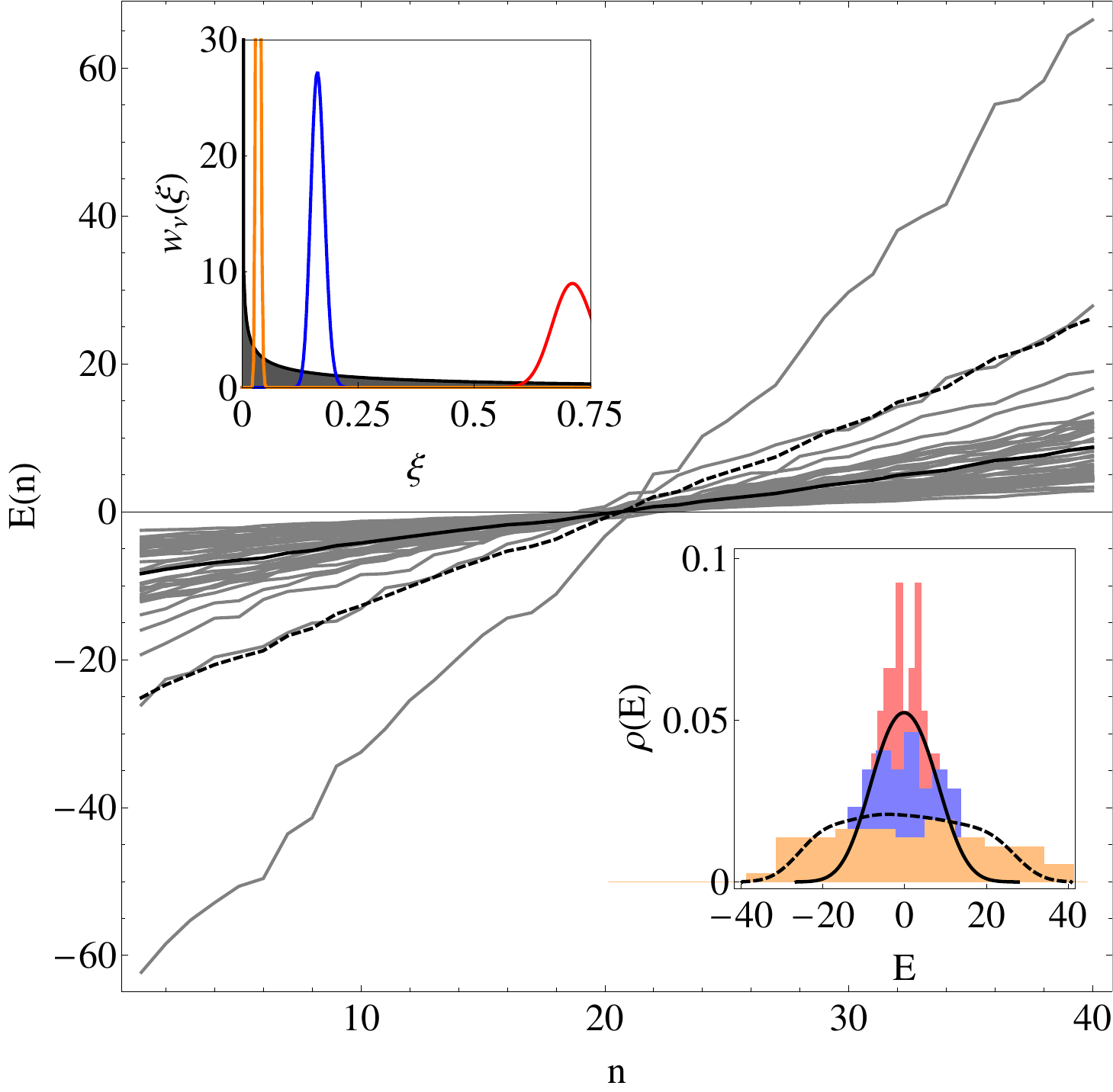}\\
        (a) Nonergodic regime ($\overline{\xi}=0.5$)
        \end{center}
        \end{minipage}
\end{minipage}
\caption{Disordered ensemble with $M$ realizations of eigenspectra with $N$ levels, using $N=M=50$, for different shapes of the initial disorder distribution: (a) a Gaussian-shaped gamma distribution $w_0(\xi)$ for $\overline{\xi}=200$, and (b) a long-tailed gamma distribution $w_0(\xi)$ for $\overline{\xi}=0.5$. (Upper left insets) Three different realizations are shown for the the convergence of the disorder distribution $w_\nu(\xi)$ (non-shaded curves) towards similar positions along the initial gamma distribution $w_0(\xi)$ (black shaded curve) for $\overline{\xi}=200$, and to very different positions for $\overline{\xi}=0.5$. (Lower right insets) Eigenvalue density histograms $\rho(E)$ of the same three realizations, which are ergodic for $\overline{\xi}=200$, and nonergodic for $\overline{\xi}=0.5$. (Main figure) Eigenvalue sequences $E^{(m)}(n)$ (continuous grey lines) for all $m=1,\ldots, M$ realizations, which are ergodic for $\overline{\xi}=200$, and nonergodic for $\overline{\xi}=0.5$. Also shown are the parameter-free and data-adaptive global behavior $\overline{E}(n)$ and $\rho(\overline{E})$ for one particular realization (dashed black line), and the ensemble average $\left< E(n) \right>$ and $\left< \rho\right>$ (continuous black lines). For the matrix dimensions $N \times N$ used in this calculation, the global level density is midway between a semicircle and a Gaussian distribution. }
\label{FigDisorderedEnsemble}
\end{figure*}

\section{A random-matrix model for nonergodic disordered ensembles}
\label{SectBohigasModel}

Let $H_G(\sigma)$ be a random matrix from GOE (with Dyson index $\beta=1$) of dimension $N \times N$, whose matrix elements are chosen independently from the Gaussian distribution $\mathcal{N}(\mu,\sigma)$, with $\mathcal{N}(0,1)$ for the diagonal elements, and $\mathcal{N}(0,1/\sqrt{2})$ for the nondiagonal elements \cite{ede05}. The ergodicity of the Gaussian ensembles can be broken by the introduction of disorder. Since one of the characteristics of a disordered system is the competition between two types of independent random variables, we can construct a nonergodic disordered random-matrix ensemble, $H(\sigma,\xi)$, by imposing an external source of randomness, $\xi$, to the fluctuations of the Gaussian matrix~\cite{boh08,boh11},
\begin{equation}
H(\sigma, \xi)=\frac{H_G(\sigma)}{\sqrt{\xi/\overline{\xi}}},
\label{EqDisorderedEnsemble1}
\end{equation}
where $\xi$ is a positive random variable chosen from a normalized probability distribution $w(\xi)$ with average $\overline{\xi}$ and variance $\sigma_\xi^2$. A demonstration of the nonergodicity of the disordered ensemble introduced through Eq.~\ref{EqDisorderedEnsemble1} is given in \cite{boh08} (Eqs. 11-14 in that reference). Although we can obtain the matrices of the disordered ensemble directly from Eq.~(\ref{EqDisorderedEnsemble1}), they may also be generated taking into account the correlations among their elements~\cite{boh08}. More specifically, it is possible to iteratively generate all the matrix elements of $H(\sigma,\xi)$, in the following way: at the $\nu$th step, a new element is sorted through the relation,
\begin{equation}
h_\nu=\frac{h_G(\sigma)}{\sqrt{\xi_\nu/\overline{\xi}}}.
\label{EqDisorderedEnsemble2}
\end{equation}
Here, $h_G$ are the $f=N(N+1)/2$ independent Gaussian matrix elements, ordered in such a way that the first $N$ ones are the diagonal elements $H_{ii}$, and the remaining ones are the rescaled off-diagonal elements $\sqrt{2} H_{ij}$. The reason for the factor $\sqrt{2}$ is explained in Ref. \cite{tos04}. Subsequent values for the disorder random variable $\xi_\nu$ are sorted recursively from a disorder distribution,
\begin{equation}
w_\nu(\xi)=\frac{w_0(\xi) \xi^{(n-1)/2} \exp \left( - \frac{\beta \xi}{\sqrt{2}\sigma \overline{\xi}} \sum_{i=1}^{\nu-1} h_i^2 \right)}{\int d \xi w_0(\xi) \xi^{(n-1)/2} \exp \left( - \frac{\beta \xi}{\sqrt{2}\sigma \overline{\xi}}  \sum_{i=1}^{\nu-1} h_i^2 \right)}.
\label{EqDisorderPotential}
\end{equation}
Fixing the set of disorder variables $\xi_1, \xi_2, \ldots, \xi_f$ during the realization of a particular matrix for the ensemble maintains the univariance of $w_\nu(\xi)$ at all time, and allows it to converge rapidly with iteration number $\nu$ to a very narrow and peaked distribution around a mean value $\overline{\xi}$, where the position of $\overline{\xi}$ depends on the shape of the initial distribution $w_0(\xi)$. Now, let us consider as a particular choice for the distribution $w_0(\xi)$ the normalized gamma distribution\footnote{Other choices of the initial distribution can also be made. Note that the nonergodic behavior depend on the width of the distribution $w(\xi)$, in such a way that for a wide distribution it is expected that averages over the ensemble of matrices will not coincide with averages over one spectrum.},
\begin{equation}
w_0(\xi)=\exp(-\xi)\xi^{\overline{\xi}-1}/\Gamma(\overline{\xi}),
\end{equation}
with  $\sigma_\xi^2=\overline{\xi}$. The mean $\overline{\xi}$ controls the behavior of the distribution $w_0(\xi)$, which can be Gaussian-like (for very large values of $\overline {\xi}$), or long-tailed (for very small values of $\overline{\xi}$). The disorder distribution $w_\nu(\xi)$ will tend to converge to similar positions for Gaussian-like $w_0(\xi)$, and to different positions for long-tailed $w_0(\xi)$, see Fig.~\ref{FigDisorderedEnsemble} (upper left insets). This can be understood through the coefficient of variation (CV), which considers standard deviation relative to the mean \cite{fos13a}, and in the present case behaves as $CV=1/\sqrt{\overline{\xi}}$, and tends to zero for large $\overline{\xi}$. This means that for very large values of $\overline{\xi}$ there will be little variation between the random initial positions for the disorder distribution, whereas for very small values of $\overline{\xi}$ all initial positions will likely be very different. Finally, the factor $(\xi/\overline{\xi})^{-1/2}$ multiplying the Gaussian matrices in Eq.~(\ref{EqDisorderedEnsemble1}) acts on the variance $\sigma^2$ of the Gaussian ensembles. Subsequent realizations of the matrix are generated using different sets of $\xi$, and the variance of each matrix depends on the width of $w_0(\xi)$, which in this way defines the ergodicity of the ensemble. This can be appreciated in Fig.~\ref{FigDisorderedEnsemble} (lower right insets), where the level density $\rho(E)$ is very similar for all realizations for $\overline{\xi}=200$, but dissimilar for different realizations for $\overline{\xi}=0.5$. Likewise, level sequences $E(n)$ evolve in similar ways for $\overline{\xi}=200$, but behave differently for $\overline{\xi}=0.5$ (main panel).
\begin{figure*}[htb!]
\begin{minipage}{\linewidth}    
        \begin{minipage}{0.5\linewidth}
        \begin{center}
        \includegraphics[width=0.97\linewidth]{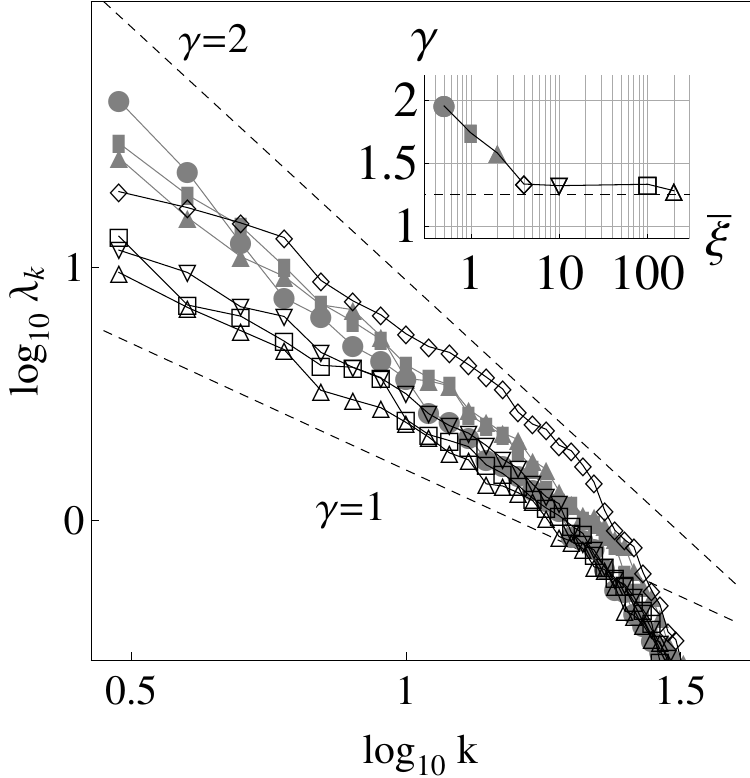}\\
        (a) Scree diagram (fluctuation part)
        \end{center}
        \end{minipage}
        \begin{minipage}{0.5\linewidth}
        \begin{center}
        \includegraphics[width=\linewidth]{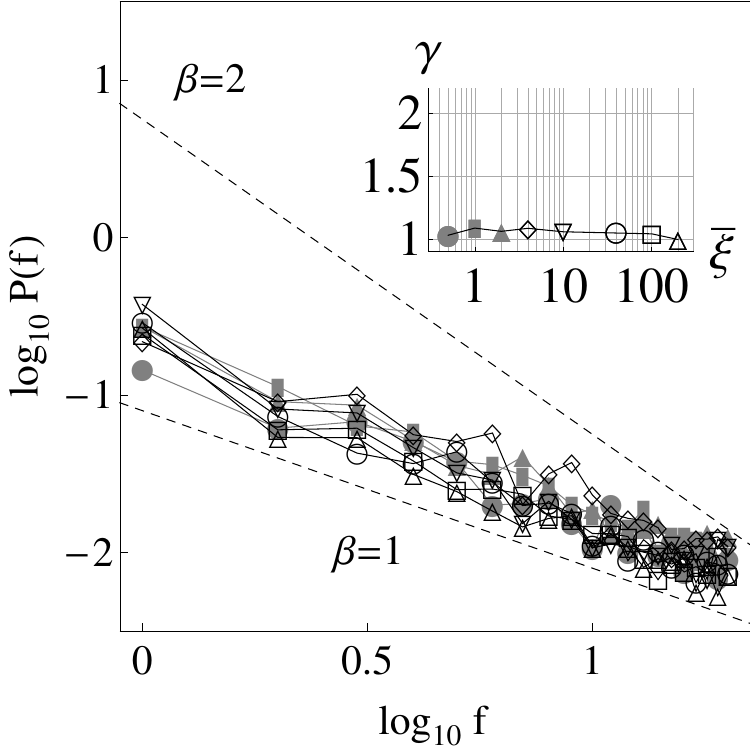}\\
        (b) Fourier power spectrum
        \end{center}
        \end{minipage}
\end{minipage}
\caption{Analysis of eigenspectra in the nonergodic regime for $\overline{\xi}=0.5,1,2$  (grey filled symbols) and in the ergodic regime $\overline{\xi}=4,10,100,200$  (black open symbols). (a) Ensemble perspective: The fluctuation part of the scree diagram $\lambda_k$ changes its scaling behavior from $\gamma=2$ (Poisson statistics) for $\overline{\xi}=0.5$ towards $\gamma < 1.5$ for $\overline{\xi} \geq 4$, compared with the numerical result $\gamma \approx 1.25$ for a very small GOE ensemble with $N=M=50$ (horizontal dashed line). (b) Individual spectrum perspective: The Fourier power spectrum $P(f)$ of the fluctuations $\widetilde{E}=E-\overline{E}$ of individual spectra results in $\beta_{ps}=1$ (GOE statistics) independently from $\overline{\xi}$.  }
\label{FigUnfolding}
\end{figure*}

\section{Normal mode analysis using Singular Value Decomposition (SVD)}
\label{SectUnfolding}

\subsection{Trend and fluctuation normal modes}

In the following, we briefly review the normal mode analysis of Refs.~\cite{fos13b,Torres17,Torres18}. Consider an ensemble of $m=1 \ldots M$ level sequences $E^{(m)}(n)$, where each sequence consists of $n=1 \ldots N$ levels, such as the ensembles with $N=M=50$ presented in Fig.~\ref{FigDisorderedEnsemble} (main panel). Each sequence constitutes one of the rows of a $M \times N$ dimensional matrix $\mathbf{X}$, which we will interprete as a multivariate time series,
\begin{equation}
\mathbf{X}=
\left(%
\begin{array}{cccc}
  E^{(1)}(1) & E^{(1)}(2) & \cdots & E^{(1)}(N) \\
  E^{(2)}(1) & E^{(2)}(2) & \cdots & E^{(2)}(N) \\
  \vdots & \vdots & \ddots & \vdots \\
  E^{(M)}(1) & E^{(M)}(2) & \cdots & E^{(M)}(N) \\
\end{array}%
\right).
\label{EqX}
\end{equation}
Singular Value Decomposition (SVD) is an exact and parameter-free matrix decomposition technique that allows us to rewrite $\mathbf{X}$ in a unique way as,
\begin{equation}
\mathbf{X} = \mathbf{U} \mathbf{\Sigma} \mathbf{V}^T =  \sum_{k=1}^r \sigma_k \vec{u}_k \vec{v}_k^T, \label{EqSVD}
\end{equation}
where $\mathbf{\Sigma}$ is an $M \times N$-dimensional matrix with only diagonal elements that are the ordered \emph{singular values} $\sigma_1 \geq \sigma_2 \geq \ldots \geq \sigma_r$, where $r \leq \mathrm{Min}[M,N]=\mathrm{rank}(\mathbf{X})$. The vectors $\vec{u}_k$ are orthonormal and they constitute the $k$th columns of the $M \times M$-dimensional matrix $\mathbf{U}$.
They are called the \emph{left-singular vectors} of $\mathbf{X}$, and they span its column space. Their physical significance will be explained further on.
The vectors $\vec{v}_k$ are orthonormal and they constitute the $k$th columns of the $N \times N$-dimensional matrix $\mathbf{V}$.
They are called the \emph{right-singular vectors} of $\mathbf{X}$, they span its row space, and therefore they constitute a basis of energy \emph{normal modes} for the ensemble.
The expression $\vec{u}_k \vec{v}_k^T \equiv \vec{u}_k \otimes \vec{v}_k$ indicates the outer product of  $\vec{u}_k$ and $\vec{v}_k$.
A set $\{\sigma_k,\vec{u}_k,\vec{v}_k \}$ is called an \emph{eigentriplet}, and completely defines the eigenmode of order $k$.
Any matrix row of $\mathbf{X}$ containing a particular eigenspectrum can be written as,
\begin{equation}
E^{(m)}(n) = \overline{E}(n)+\widetilde{E}(n)=\sum_{k=1}^r \sigma_k U_{mk} \vec{v}^T_k(n),
\label{EqDecomposition}
\end{equation}
where $\lambda_k=\sigma_k^2$ can be interpreted as \emph{partial variances} that indicate how much a specific normal mode $\vec{v}_k$ contributes to the total variance of the ensemble, and the matrix elements $U_{mk}$ serve as coefficients that express a particular level sequence exactly as a weighted sum of normal modes. The normal modes $\vec{v}_k$ with $k=1, \ldots, n_T$ that determine the global spectral properties $\overline{E}$ of a particular spectrum 
behave monotonously and can easily be distinguished by their large partial variances $\lambda_k$ that are orders of magnitude larger than the remaining $\lambda_k$ with $k=n_T+1,\ldots, r$ associated to the oscillating normal modes of the fluctuations $\widetilde{E}$ \cite{fos13b}. In the present calculation, we find one global mode  ($n_T=1$) for the matrix ensembles in the ergodic regime, and two global modes ($n_T=2$) in the nonergodic regime. In Fig.~\ref{FigDisorderedEnsemble} (main panels), the global behavior $\overline{E}$ is shown for a particular level sequence, in comparison with the ensemble mean $\left< E \right>$. The global level density is easily calculated as the histogram $\rho(\overline{E})$ of the global behavior $\overline{E}$ and is compared with the ensemble average $\left< \rho \right>$ (lower right insets). It can be appreciated that in the ergodic regime the ensemble average $\left< E \right>$ or $\left< \rho \right>$ is representative for an individual spectral average $\overline{E}$ or $\rho(\overline{E})$, whereas in the nonergodic regime the ensemble mean is not representative.
\begin{figure}[htb!]
\begin{center}
\includegraphics[width=0.65\linewidth]{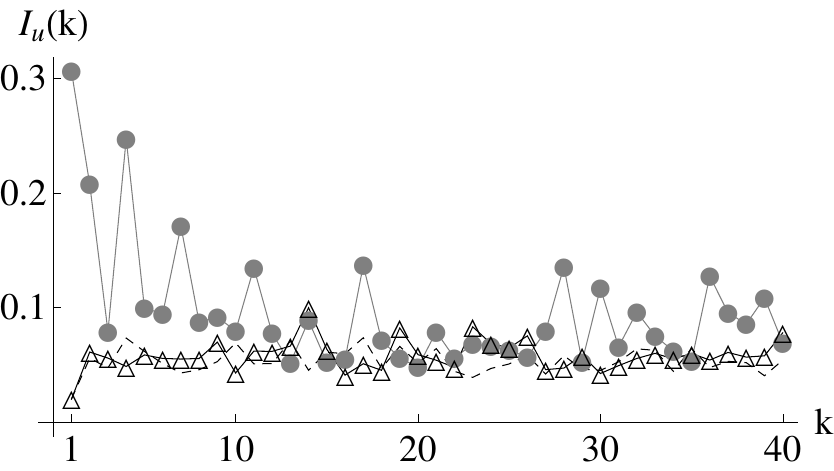}\\
\end{center}
\caption{Inverse Participation Ratio $I_u(k)$ of the left-singular vectors $\vec{u}_k$. In the ergodic regime ($\overline{\xi}=200$, black open triangles), $I_u(k)$ is low for most $\vec{u}_k$ indicating that these vectors are influenced by most spectra and thus are representative for the whole ensemble. In the nonergodic regime ($\overline{\xi}=0.5$, gray filled circles), peaks of high $I_u(k)$ indicate that these vectors are influenced by one or only few spectra and thus are not representative for the whole ensemble. In comparison, results for $I_u(k)$ are shown for an ensemble of GOE spectra of the same dimensions $N=M=50$ (black dashed lines). }
\label{FigInverseParticipationRatio}
\end{figure}

\subsection{Fluctuation measures}

In Refs.~\cite{fos13b,Torres17,Torres18}, we applied the above normal mode analysis to ergodic Gaussian ensembles. On the one hand, 
the fluctuation part $k=n_T+1, \ldots, r$ of the \emph{scree diagram} of ordered partial variances behaves as a power law, \begin{equation}
\lambda_k \propto 1/k^\gamma,
\label{EqScreePowerLaw}
\end{equation}
which gives the \emph{ensemble estimate} of the spectral rigidity in terms of how the normal modes $\vec{v}_k$ common to all eigenspectra scale. On the other hand,
the Fourier power spectrum of the fluctuations $\widetilde{E}(n)$ of the individual spectra also follows a power law,
\begin{equation}
P(f) \propto 1/f^{\beta_{ps}}.
\label{EqFourierPowerLaw}
\end{equation}
which is the \emph{spectrum estimate} of the spectral rigidity.
For ergodic ensembles, it resulted that the spectral exponents of both estimates are equal \cite{fos13b}. The value $\beta_{ps}=\gamma=1$ characterizes correlated spectra of the GOE type because of level repulsion. The value $\beta_{ps}=\gamma=2$ reflects the Poissonian statistics of noncorrelated levels in the absence of level repulsion. \\

In the following, we analyze eigenspectra of the disordered matrix ensemble of Sect.~\ref{SectBohigasModel}. In Fig.~\ref{FigUnfolding} (panel (a)), the fluctuation part of the scree diagram $\lambda_k$ is shown for different values of $\overline{\xi}$. Apart from a tail of nonsignificant $\lambda_k$ for high-order modes when the basis becomes over-complete~\cite{fos13b}, the power-law behaviour of Eq.~(\ref{EqScreePowerLaw}) is observed for all realizations, and the value of the spectral exponent $\gamma$ changes in function of $\overline{\xi}$ (see inset). For $\overline{\xi}=0.5$, in the nonergodic regime, we find $\gamma=2$ corresponding to Poisson statistics. The spectral exponent $\gamma$ drops quickly for increasing $\overline{\xi}$, reflecting a rapid decrease of intensity of nonergodicity, as will be explained in the next subsection. For $\overline{\xi} \geq 4$, there is a further low approach to the ergodic limit, in correspondence with the results of Ref.~\cite{boh08} for the fluctuation measure $\Sigma^2$, obtained after a traditional ensemble unfolding. The expected value of $\gamma=1$ for the ergodic limit is never obtained because of the small ensemble dimensions used in the present calculations $N=M=50$, where the power law can be followed only over a very limited range of less than one order of magnitude. For a GOE ensemble with the same limited dimensions $N$ and $M$, the numerical result $\gamma \approx 1.25$ is obtained.  \\

On the other hand, in Fig.~\ref{FigUnfolding} (panel (b)), the Fourier power spectrum of the fluctuations $\widetilde{E}$ of individual eigenspectra follows the power law of Eq.~(\ref{EqFourierPowerLaw}) with spectral exponent $\beta_{ps}=1$, indicating GOE statistics, as indeed expected if the unfolding is carried out appropriately \cite{boh08}. Moreover, the value for spectral exponent $\beta_{ps}$ is independent from $\overline{\xi}$. The spectral exponent $\gamma$ approaches the value for $\beta_{ps}$ for larger values of the control parameter $\overline{\xi}$ when the ergodic limit is approached. Based on the above, we propose that the difference between the spectral exponents $\beta_{ps}$ and $\gamma$ can serve as a measure of nonergodicity.

\subsection{Inverse participation ratio and breakdown of the normal-mode basis}

Now, we explain the value of the spectral exponent $\gamma$. The component $m$ of a given left-singular vector $\vec{u}_k$ relates to the contribution of spectrum $m$ to that vector. Hence, the distribution of the components contains information about the number of spectra contributing to a specific left-singular vector. In order to distinguish between one vector with approximately equal components and another with a small number of large components, one can define the inverse participation ratio for a vector $\vec{u}_k$~\cite{Bell70,Thou74,ple99,ple02},
\begin{equation}
I_u(k) \equiv \sum_{m=1}^M \left[ \vec{u}_k(m) \right]^4.
\end{equation}
The physical meaning of $I_u(k)$ can be illustrated by two limiting cases, (i) an eigenvector with identical components $\vec{u}_k(m)=1/\sqrt{M}$ has $I_u(k)=1/M$, whereas (ii) an eigenvector with one component $\vec{u}_k(m)=1$ and all the others zero has $I_u(k)=1$. Therefore, $I_u(k)$ is related to the reciprocal of the number of eigenvector components significantly different from zero.\\

In Fig.~\ref{FigInverseParticipationRatio}, we can see that in the nonergodic regime for $\overline{\xi}=0.5$ almost half of the vectors $\vec{u}_k$ has a very high inverse participation ratio $I_u(k)$, indicating that one or only few spectra contribute to the eigentriplet  $\{\sigma_k,\vec{u}_k,\vec{v}_k \}$, so that this eigenmode is not representative for the whole ensemble. The scree diagram is thus composed of many noncorrelated partial variances, resulting in Poissonian statistics. In the ergodic regime, for $\overline{\xi}=200$, inverse participation ratios $I_u(k)$ are small, indicating that most if not all spectra contribute to the eigentriplets, which are thus representative for the whole ensemble. The scree diagram is composed of fully correlated partial variances and results in GOE statistics. The few moderate peaks that appear in $I_u(k)$ for $\overline{\xi}=200$ with respect to the results for a GOE ensemble of the same dimensions $N=M=50$ indicates that the fully ergodic limit has not yet been reached.\\

In this context, nonergodicity can be understood as a breakdown of the common normal-mode basis of the ensemble, not only at the large energy scale of the global spectral behavior $\overline{E}$, see Fig.~\ref{FigDisorderedEnsemble} (main panels), but also at the small scale of the local fluctuations $\widetilde{E}$, see Fig.~\ref{FigUnfolding} (panel (a)). In this way, the inverse participation ratio $I_u(k)$ serves as a measure for nonergodicity.

\subsection{Rescaling}

By applying the normal mode analysis presented here, the spectra are not unfolded, i.e., they are not rescaled but only detrended. If the purpose is to calculate traditional fluctuation measures such as the short-range nearest-neighbor spacing distribution (NNSD) or the long-range $\Sigma^2$ or $\Delta_3$ measures, which require an explicit normalization of the fluctuations \cite{meh91}, a data-adaptive unfolding can be performed using the fluctuation normal modes~\cite{tor14,fos14,fos15}. Time-series based fluctuation measures such as the scree diagram $\lambda_k$, or the Fourier power spectrum $P(f)$, absorb the scale of the fluctuations in the offset of the power law of Eqs.~(\ref{EqScreePowerLaw}) and (\ref{EqFourierPowerLaw}), whereas the statistics of the fluctuations is codified in the spectral exponents $\gamma$ and $\beta_{ps}$.\\

This can be illustrated with Fig.~\ref{FigUnfolding} (panel (a)), where the offset of the fluctuation part of the scree diagram varies over almost a whole order of magnitude. For very large values of $\overline{\xi}$, the factor that determines the variance of the ensemble $(\xi/\overline{\xi})^{-1/2} \rightarrow 1$, and the variance of the disordered ensemble $H(\sigma,\xi)$ tends to remain unchanged with respect to the initial Gaussian ensemble $H(\sigma)$ (curves with black open symbols). For very small values of $\overline{\xi}$, the factor $(\xi/\overline{\xi})^{-1/2}$ can become very large because of the divergence of $w_0(\xi)$ near $\xi=0$, and the variance of the disordered ensemble is enhanced (curves with grey filled symbols). The case $\overline{\xi}=4$ is intermediate between these two regimes.

\section{Conclusions}
\label{SectConclusion}

In the present contribution, we applied the SVD method to a disordered random-matrix model, that allows to fine tune the intensity of nonergodicity. In this way, we calculated ensemble-averaged and spectrum-averaged statistics in a parameter-free and consistent way, without performing any unfolding of the spectra, by using a data-adaptive basis of normal modes. In this context, nonergodicity was explained as the breakdown of the common normal mode basis, so that the inverse participation ratio, and the difference between the spectrum-averaged and the ensemble-averaged statistics, served as measures for the intensity of nonergodicity. Results obtained suggest that SVD could be a robust tool for characterizing spectra of systems in which nonergodicity may play an important
role such as network spectra.

\section*{Acknowledgements}

Financial funding for this work was supplied by the Direcci\'on General de Asuntos del Personal Acad\'emico (DGAPA) from the Universidad Nacional Aut\'onoma de Mexico (UNAM) with grant IA102619. We are thankful for the Newton Advanced Fellowship awarded to RF by the Academy of Medical Sciences through the UK Governments Newton Fund program. The authors wishes to thank J.~C. L\'opez-Vieyra and V. Vel\'azquez for fruitful discussions.


\begin{thebibliography}{00}

\bibitem{meh91} M. L. Mehta, Random matrices (Acad. Press, New York, 1991), 2nd ed.
\bibitem{haa10} F. Haake, \emph{Quantum signatures of chaos} (Springer, Heidelberg, 2010), 3rd ed.
\bibitem{ben01} L. Benet, T. Rupp and H. A. Weidenm\"{u}ller, Phys. Rev. Lett. \textbf{87}, 010601 (2001).
\bibitem{asa01} T. Asaga, L. Benet, T. Rupp and H. A. Weidenm\"{u}ller, Europhys. Lett. \textbf{56}, 340 (2001).
\bibitem{jac01} A. D. Jackson, C. Mejia-Monasterio, T. Rupp, M. Saltzer and T. Wilke, Nucl. Phys. A \textbf{687}, 405 (2001).
\bibitem{mal07} C. Male, G. Le Ca\"{e}r and R. Delannay, Phys. Rev. E \textbf{76}, 042101 (2007).
\bibitem{rel08} A. Rela\~{n}o, L. Mu\~{n}oz, J. Retamosa, E. Faleiro and R. A. Molina, Phys. Rev. E \textbf{77}, 031103 (2008).
\bibitem{fre70} J. B. French and S. S. M. Wong, Phys. Lett. B \textbf{33}, 449 (1970); O. Bohigas and J. Flores, ibid. \textbf{34}, 261 (1971); J. B. French and S. S. M. Wong, ibid. \textbf{35}, 383 (1971).
\bibitem{flo01} J. Flores, M. Horoi, M. M\"{u}ller and T. H. Seligman, Phys. Rev. E \textbf{63}, 026204 (2001).
\bibitem{mon75} K. K. Mon and J. B. French, Ann. Phys. \textbf{95}, 90 (1975); L. Benet and H. A. Weidenm\"{u}ller, J. Phys. A: Math. Gen. \textbf{36}, 3569 (2003).
\bibitem{ciz94} P. Cizeau and J. P. Bouchaud, Phys. Rev. E \textbf{50}, 1810 (1994).
\bibitem{guh90} T. Guhr and H. A. Weidenm\"{u}ller, Ann. Phys. (N.Y.) \textbf{199}, 412 (1990); M. S. Hussein and M. P. Pato, Phys. Rev. Lett. \textbf{70}, 1089 (1993).
\bibitem{bir07} G. Biroli, J. P. Bouchaud and M. Potters, Acta Phys. Pol. B \textbf{38}, 4009 (2007).
\bibitem{boh08} O. Bohigas, J. X. de Carvalho and M. P. Pato, Phys. Rev. E \textbf{77}, 011122 (2008).
\bibitem{boh11} O. Bohigas and M. P. Pato, Phys. Rev. \textbf{E}, 031121 (2011).
\bibitem{tos04} F. Toscano, R. O. Vallejos and C. Tsallis, Phys. Rev. E \textbf{69}, 066131 (2004).
\bibitem{ber04} A. C. Bertuola, O. Bohigas and M. P. Pato, Phys. Rev. E \textbf{70}, 065102(R) (2004).
\bibitem{bro81} T. A. Brody, J. Flores, J. B. French, P. A. Mello, A. Pandey and S. S. M. Wong, Rev. Mod. Phys. \textbf{53}, 385 (1981).
\bibitem{gom02} J. M. G. G\'omez, R. A. Molina, A. Rela\~no and J. Retamosa, Phys. Rev. E \textbf{66}, 036209 (2002).
\bibitem{abu12} S. M. Abuelenin and A. Y. Abul-Magd, Pocedia CS \textbf{12}, 69 (2012).
\bibitem{she18} S. M. Abuelenin, Phys. A \textbf{492}, 564 (2018).
\bibitem{fos13b} R. Fossion, G. Torres Vargas and J. C. L\'opez Vieyra, Phys. Rev. E \textbf{88}, 060902(R) (2013).
\bibitem{Torres17} G. Torres-Vargas, R. Fossion, C. Tapia-Ignacio and J. C. L\'opez-Vieyra, Phys. Rev. E, \textbf{96}, 012110 (2017).
\bibitem{Torres18} G. Torres-Vargas, J. A. M\'endez-Berm\'udez, J. C. L\'opez-Vieyra and R. Fossion, Phys. Rev. E, \textbf{98}, 022110 (2018).
\bibitem{fos15} R. Fossion, G. Torres-Vargas, V. Vel\'azquez and J. C. L\'opez-Vieyra, J. Phys.: Conf. Ser. \textbf{578}, 012013 (2015).
\bibitem{fos13a} R. Fossion, D. A. Hartas\'anchez, O. Resendis-Antonio and A. Frank, Front. Biol. \textbf{8}, 247 (2013).
\bibitem{ede05} A. Edelman and N. Raj Rao, Acta Num. \textbf{14}, 233 (2005).
\bibitem{Bell70} R.J. Bell and P. Dean, Disc. Faraday Soc. \textbf{50}, 55-61 (1970).
\bibitem{Thou74} D. J. Thouless, Phys. Reps. \textbf{13}, 93-142 (1974).
\bibitem{ple99} V. Plerou, P. Gopikrishnan, L.A. Nunes Amaral and H. Eugene Stanley, Phys. Rev. Lett. \textbf{83}, 1471 (1999).
\bibitem{ple02} V. Plerou, P. Gopikrishnan, B. Rosenow, L.A. Nunes Amaral, T. Guhr and H. Eugene Stanley, Phys. Rev. E \textbf{65}, 066126 (2002).
\bibitem{tor14} G. Torres Vargas, R. Fossion, V. Vel\'azquez and J. C. L\'opez Vieyra, J. Phys.: Conf. Ser. \textbf{492}, 012011 (2014).
\bibitem{fos14} R. Fossion, AIP Conference Proceedings \textbf{1575}, 89 (2014).
\end{thebibliography}
\end{document}